\documentclass[journal=jacsat,manuscript=article]{achemso}

\usepackage[version=3]{mhchem} 
\usepackage[utf8]{inputenc}
\usepackage{amsmath}
\usepackage{siunitx}
\usepackage{graphicx}

\usepackage[hidelinks]{hyperref} 
\usepackage[nameinlink]{cleveref}

\author{Yufan Li}
\affiliation[QN]
{Department of Quantum Nanoscience, Kavli Institute of Nanoscience, Delft University of Technology, Delft, The Netherlands}
\alsoaffiliation[3me]
{Department of Precision and Microsystems Engineering, Faculty of Mechanical, Maritime and Materials Engineering, Delft University of Technology, Delft, The Netherlands}

\author{Fabian A. Gerritsma}
\affiliation[QN]
{Department of Quantum Nanoscience, Kavli Institute of Nanoscience, Delft University of Technology, Delft, The Netherlands}

\author{Samer Kurdi}
\affiliation[QN]
{Department of Quantum Nanoscience, Kavli Institute of Nanoscience, Delft University of Technology, Delft, The Netherlands}

\author{Nina Codreanu}
\affiliation[Qutech]
{QuTech and Kavli Institute of Nanoscience, Delft University of Technology, Delft, The Netherlands}

\author{Simon Gr\"{o}blacher}
\affiliation[QN]
{Department of Quantum Nanoscience, Kavli Institute of Nanoscience, Delft University of Technology, Delft, The Netherlands}

\author{Ronald Hanson}
\affiliation[Qutech]
{QuTech and Kavli Institute of Nanoscience, Delft University of Technology, Delft, The Netherlands}

\author{Richard Norte}
\affiliation[QN]
{Department of Quantum Nanoscience, Kavli Institute of Nanoscience, Delft University of Technology, Delft, The Netherlands}
\alsoaffiliation[3me]
{Department of Precision and Microsystems Engineering, Faculty of Mechanical, Maritime and Materials Engineering, Delft University of Technology, Delft, The Netherlands}

\author{Toeno van der Sar}
\affiliation[QN]
{Department of Quantum Nanoscience, Kavli Institute of Nanoscience, Delft University of Technology, Delft, The Netherlands}
\email{t.vandersar@tudelft.nl}

\title[An \textsf{achemso} demo]
  {A Fiber-coupled Scanning Magnetometer with Nitrogen-Vacancy Spins in a Diamond Nanobeam}

\abbreviations{IR,NMR,UV}
\keywords{American Chemical Society, \LaTeX}

\begin{document}







\begin{abstract}
  Magnetic imaging with nitrogen-vacancy (NV) spins in diamond is becoming an established tool for studying nanoscale physics in condensed matter systems. However, the optical access required for NV spin readout remains an important hurdle for operation in challenging environments such as millikelvin cryostats or biological systems. Here, we demonstrate a scanning-NV sensor consisting of a diamond nanobeam that is optically coupled to a tapered optical fiber. This nanobeam sensor combines a natural scanning-probe geometry with high-efficiency through-fiber optical excitation and readout of the NV spins. We demonstrate through-fiber optically interrogated electron spin resonance and proof-of-principle magnetometry operation by imaging spin waves in an yttrium-iron-garnet thin film. Our scanning-nanobeam sensor can be combined with nanophotonic structuring to control the light-matter interaction strength, and has potential for applications that benefit from all-fiber sensor access such as millikelvin systems.
\end{abstract}

\section{Introduction}

The nitrogen-vacancy (NV) lattice defect in diamond has emerged as a powerful magnetic-field sensor. High-fidelity microwave control and optical readout of the NV spin \cite{Schirhagl2014Nitrogen-vacancyBiology,Rondin2014MagnetometryDiamond,Doherty2013TheDiamond} over a wide range of conditions has enabled applications in condensed matter physics \cite{Casola2018ProbingDiamond}, chemistry \cite{Mochalin2012TheNanodiamonds}, biology \cite{LeSage2013OpticalCells,Barry2016OpticalDiamond} and geoscience \cite{Glenn2017Micrometer-scaleMicroscope}. In particular, scanning-probe magnetometry based on individual NV spins in diamond nanotips has provided imaging of spins and currents in materials with spatial resolution down to $\sim\SI{50}{\nm}$ \cite{Maletinsky2012ACentres,Gross2017Real-spaceMagnetometer,Simon2021DirectionalWaves}.

An important challenge for the application of scanning-probe NV magnetometry in advanced environments such as millikelvin cryostats is the required optical access to the NV spins. Free-space optical access leads to additional heat load and increased complexity of cryostat design. A potential way to preclude the need for free-space optical access is to realize fiber-based scanning-NV sensors \cite{Fedotov2014Fiber-opticImaging,Chatzidrosos2021FiberizedMagnetometers}. Here, we demonstrate a new scanning-NV sensor based on a tapered diamond nanobeam that is optically coupled to, and manipulated with a tapered optical fiber. Such fiber-based NV nanobeam sensors could facilitate implementation in low-temperature setups, while benefiting from the potentially near-perfect optical coupling efficiency between fiber and nanobeam \cite{Tiecke2015EfficientDevices,Burek2017Fiber-coupledInterface,Groblacher2013HighlyCavity}. Moreover, nanobeams are excellently suited for nanophotonic structuring \cite{Burek2014HighDiamond,Mouradian2017RectangularDiamond}, which could enable high-efficiency, resonant optical addressing of embedded NV centers or other group-IV color centers\cite{bradac2019quantum} by incorporating photonic crystals. 

\begin{figure}[hbt]
    \centering
    \includegraphics[width=\textwidth]{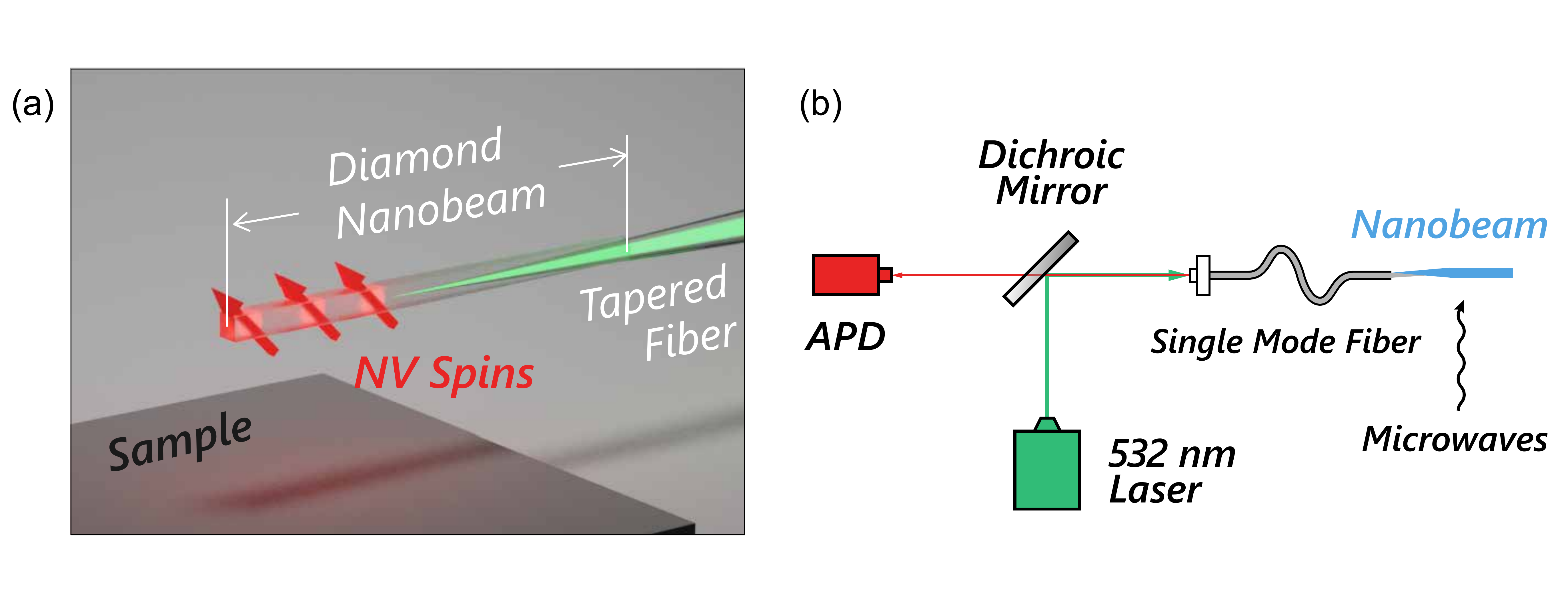}
    \caption{Magnetometry based on nitrogen-vacancy spins in fiber-coupled diamond nanobeams. (a) Schematic illustration of the technique: A diamond nanobeam with an ensemble of NV spins is optically coupled to a tapered fiber. The fiber both guides the excitation laser to the NVs and collects the NV photoluminescence, resulting in a scanning-probe magnetometer that does not require free-space optical access. Scanning the nanobeam parallel to the sample surface enables magnetic imaging with high resolution in the direction perpendicular to the beam axis. (b) Simplified scheme of the optics used to excite and read out the NV photoluminescence. A 532 nm laser excitation is coupled into the fiber that delivers the light to the NV centers in the diamond nanobeam. The resulting NV photoluminescence is collected through the same fiber, separated from the excitation light by a dichroic mirror, and detected by an avalanche photodiode (APD). }
    \label{fig:schematic}
\end{figure}

We fabricate the diamond nanobeams using nanofabrication recipes developed in Refs. 
\citenum{Mouradian2017RectangularDiamond,Khanaliloo2015High-Q/VEtching,Wan2020Large-scaleCircuits,max2021thesis}. The key advance we present here is the ability to break off and attach individual tapered diamond nanobeams to nanoscale-tapered optical fibers and use these nanobeam sensors for scanning NV magnetometry (\cref{fig:schematic}(a)). We break a beam off the bulk diamond by pushing on it with the fiber, after which the beam and the fiber remain attached, presumably by van der Waals forces. We found it crucial to create long ($\sim\SI{40}{\micro\meter}$) nanobeams to provide a sufficient lever arm that prevents damaging the glass fiber tip. The use of $\sim\SI{70}{\nano\meter}$ wide bridges between  the nanobeam and bulk diamond (\cref{fig:fab}(b)) enables easy breaking. As a proof of principle, we demonstrate through-fiber optical interrogation of an NV-nanobeam sensor, characterize its photon collection efficiency, and demonstrate its imaging capability by visualizing spin waves in  a thin film of yttrium-iron-garnet (YIG) \cite{Simon2021DirectionalWaves,Bertelli2020MagneticInsulator}. 

\section{Results and Discussion}

\begin{figure}[ht!]
    \centering
    \includegraphics[width=\textwidth]{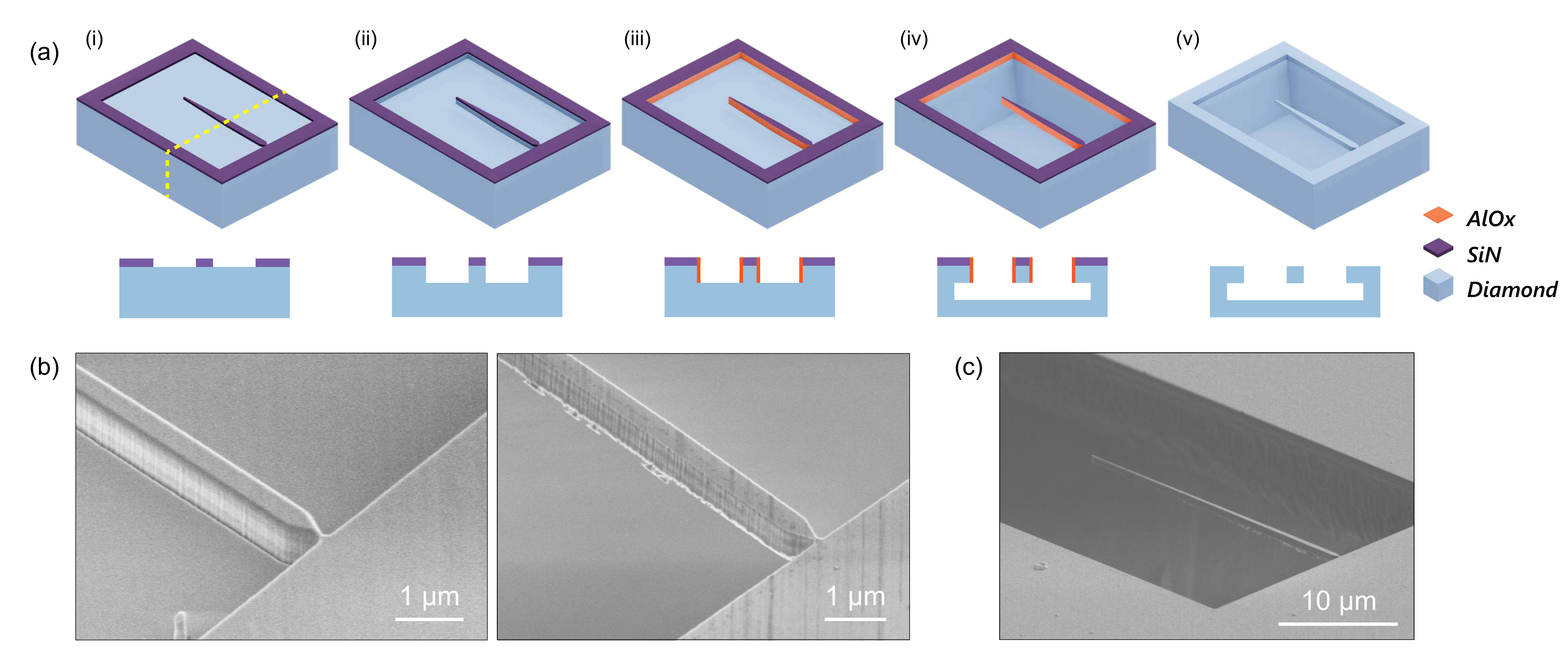}
    \caption{Fabrication of large-aspect-ratio diamond nanobeams that enable breaking-off and fiber coupling. (a) Fabrication flow \cite{Khanaliloo2015High-Q/VEtching}. i) A $\mathrm{Si_3N_4}$ hard mask ($\sim\SI{200}{\nm}$ thickness, purple) is fabricated on a diamond substrate (blue) using electron-beam lithography and anisotropic $\mathrm{CHF_3/O_2}$ reactive ion etching (RIE). ii) An anisotropic $\mathrm{O_2}$ RIE process defines the nanobeam sidewalls in the diamond. iii) A $\sim\SI{20}{\nm}$ layer of $\mathrm{Al_2O_3}$ (orange) is grown by atomic layer deposition (ALD) to protect the nanobeam sidewalls during the subsequent undercut step. An anisotropic $\mathrm{BCl_3/Cl_2}$ RIE step removes the $\mathrm{Al_2O_3}$ on the horizontal surfaces. iv) An isotropic $\mathrm{O_2}$ RIE process undercuts the diamond nanobeam. v) Removal of all masks with hydrofluoric (HF) acid. The schematics beneath each panel show corresponding cross-sectional views, marked with the yellow dashed line in panel (i). (b) Scanning electron microscope (SEM) images of representative nanobeams during fabrication (zoomed in to show the connection point), taken at stages illustrated in panels (ii) (left) and (v) (right) in (a). (c) SEM image of a $\SI{40}{\micro\meter}$ long diamond nanobeam after the fabrication. }
    \label{fig:fab}
\end{figure}

Efficient optical coupling to the fiber requires tapered diamond nanobeams with nanoscale widths \cite{Burek2017Fiber-coupledInterface}. We fabricate these beams out of a single-crystal diamond chip using the procedure demonstrated in Ref. \citenum{max2021thesis} (\cref{fig:fab}). We first deposit a $\SI{200}{\nm}$-thick $\mathrm{Si_3N_4}$ mask onto the diamond using plasma-enhanced chemical vapor deposition (PECVD). We then use e-beam lithography and reactive ion etching (RIE) with a $\mathrm{CHF_3/O_2}$ plasma to pattern the beams and their holding bars (``tethers") on the $\mathrm{Si_3N_4}$ hard mask. An anisotropic, inductively-coupled plasma (ICP) RIE with O$_2$ transfers the patterns from the hard mask to the diamond substrate (\cref{fig:fab}(b)). Using atomic layer deposition (ALD) to deposit $\SI{20}{nm}$ of $\mathrm{Al_2O_3}$ \cite{Mouradian2017RectangularDiamond}, we create a conformal layer that protects the vertical sidewalls during the subsequent undercut. An anisotropic ICP-RIE with $\mathrm{BCl_3/Cl_2}$ removes the $\mathrm{Al_2O_3}$ from the horizontal diamond surfaces while leaving the vertical beam sidewalls protected. Finally, we undercut the nanobeams with a quasi-isotropic $\mathrm{O_2}$ ICP-RIE and remove the masks with hydrofluoric acid (HF), leaving free-hanging diamond nanobeams (\cref{fig:fab}(c)).

\begin{figure}[ht!]
    \centering
    \includegraphics[width=\textwidth]{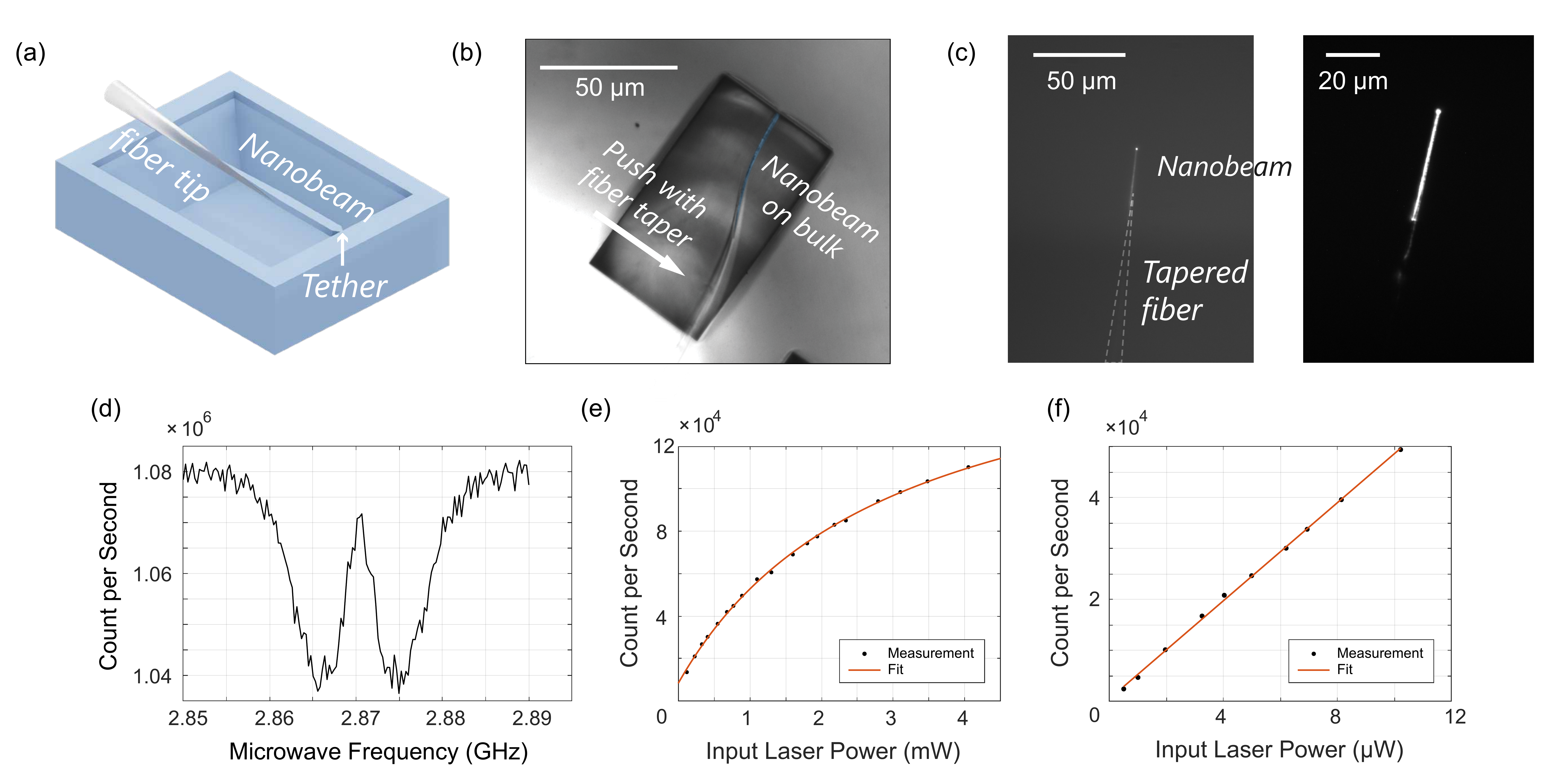}
    \caption{Assembly and characterization of a fiber-coupled diamond nanobeam sensor. (a) Schematic illustration  of a tapered fiber brought into contact with a nanobeam. (b) Microscope image of the fiber pushing sideways on the nanobeam (artificially colored in blue) to break the $\sim\SI{70}{\nm}$-wide connection to the diamond chip. Both beam and fiber bend strongly before breaking. (c) After breaking off, the beam sticks to the fiber (outlined with the dashed line). Through-fiber green-laser excitation causes the bright NV photoluminescence visible in this camera image. The image is taken with a $\SI{650}{\nm}$ long-pass filter to block the excitation light. Right panel is a zoom-in view and with higher excitation power, highlighting the NV photoluminescence in the beam. (d) NV electron spin resonance (ESR) measurement at zero magnetic field, measured using the setup depicted in \cref{fig:schematic}(b). Optical excitation power: 30 nW. (e) NV photoluminescence vs optical excitation power, fitted with \cref{eq:saturation} (orange curve). The count rate on the $y$-axis is after neutral-density (ND) filtering by a factor of $6\times 10^6$ (so that the measured photoluminescence rate stays within the measurement range of the APD). (f) Characterization of fiber autoluminescence in our experiment. Linear fit gives an autoluminescence rate of $\Gamma_\mathrm{fiber}=\SI{4.8e3}{\per\second\per\micro\watt}.$}
    \label{fig:coupling}
\end{figure}

To couple the nanobeams to a tapered optical fiber (S630-HP, tapered by HF pulling, see methods), we mount the nanobeam chip on a 3-axis slip-stick positioner (Mechonics MX-35). Monitoring through a microscope objective (Mitutoyo M Plan Apo HR 50$\times$), we push the fiber against the nanobeam by moving the stage perpendicularly to the beam until the connection point breaks and the nanobeam sticks to the fiber. The sticking is presumably due to van der Waals force. The process and end result are illustrated in \cref{fig:coupling}.

Compared to similar strategies of picking up nanophotonic structures made from other materials, such as Si or SiN \cite{Thompson2013CouplingCavity,marinkovic2021hybrid,Magrini2018Near-fieldCavity}, the main challenge lies in the significantly larger yield strength of diamond compared to glass\cite{ruoff1979yield}. Also, single crystal diamond on the nanoscale is known to exhibit large elastic deformation before fracturing when pressure is applied \cite{Banerjee2018UltralargeDiamond}, as we also observe while pushing on the beam with the fiber in \cref{fig:coupling}(b). The resulting abrupt motion when the beam breaks makes it challenging to stick the beam to the optical fiber. To overcome this challenge, we found it crucial to design beams that are at least $\SI{30}{\micro\meter}$ in length. Furthermore, we minimize the width of the tether by fabricating an array of devices with varying tether widths, and use the beams with the thinnest tethers that survived the fabrication process. Furthermore, the area of the open region around the beam should be large enough (in our design $\SI{70}{\micro\meter}\times\SI{40}{\micro\meter}$) to allow for the beam displacement during the breaking process. With these design implementations, we are able to apply a large enough torque on the tether to break the nanobeam off the bulk with a tapered optical fiber, and couple the beam to the fiber in the same process. We find tether widths of 60-80 nm to be optimal, where around 60\% of the beams remain attached to the bulk after the undercut and subsequent acid cleaning, and can be picked up by the tapered fiber with a success rate of 40\% (four out of ten beams). 


We demonstrate through-fiber optical excitation and readout of an ensemble of NV centers in a diamond nanobeam using the setup depicted in Fig. 1(b). Our nanobeams (fabricated using Element-six DNV-B14 diamond) have an estimated NV concentration of $\SI{4.5}{ppm}$ \cite{ElementSix2021DNVDatasheet}, corresponding to $N\approx4.5\times 10^6$ NVs per nanobeam ($\SI{40}{\micro\meter}$ long, maximum cross section $0.5\times\SI{0.5}{\micro\meter}^2$ and tapered down to $\sim0.1\times\SI{0.5}{\micro\meter}^2$ over $\SI{30}{\micro\meter}$ length). We apply microwaves (Windfreak SynthHD) through a co-planar waveguide to drive the electron-spin resonance (ESR) of the NV centers. \Cref{fig:coupling}(d) shows a characteristic ESR spectrum measured through-fiber from our device, where the dips result from the microwave-driven transition between NV spin states $|m_S = 0\rangle \rightarrow |m_S = \pm 1\rangle$. Due to the high NV density, we record the ESR signal with only $\SI{30}{\nano\watt}$ of excitation power. Comparing the photoluminescence measured in \cref{fig:coupling}(d) to a control measurement of the fiber autoluminescence (\cref{fig:coupling}(f)) shows that the signal is dominated by the NV photoluminescence (signal to background ratio: $8\times 10^3$) because of the high NV concentration in the nanobeam. By normalizing the photon count in \cref{fig:coupling}(d) to the total number of NVs $N$, we estimate the collected photoluminescence rate of a single NV center to be $\Gamma_\mathrm{NV}/N=\SI{8.1}{\per\second\per\micro\watt}\ll\Gamma_\mathrm{fiber}$. Therefore in order to achieve efficient single NV readout where $\Gamma_\mathrm{NV}/N\sim\Gamma_\mathrm{fiber}$, further effort is needed on both reducing the fiber autoluminescence and increasing the NV photon collection efficiency.

To estimate the photon coupling efficiency of the nanobeam-fiber interface $\eta_\mathrm{nf}$, we use two approaches. In the first, we characterize the saturation of the NV photoluminescence as a function of the optical excitation power $P$. Assuming a simple two-level model for the NV photodynamics, the NV photoluminescence is limited by the NV's spontaneous emission rate $\gamma=$1/(13 ns) \cite{Manson2006Nitrogen-vacancyDynamics}. As such, the photon count rate $\Gamma$ detected by our avalanche photodiode (APD, Laser Component COUNT-500N-FC)  can be described by:

\begin{equation}
\label{eq:saturation}
    \Gamma = \eta N\gamma\frac{P}{P+P_\mathrm{sat}} + \Gamma_\mathrm{dark}.
\end{equation}

Here, $\eta$ is the fraction of total number of photons emitted by the NVs that is detected by our APD, $P_\mathrm{sat}$ is the optical saturation power \cite{Dreau2011AvoidingSensitivity}, and $\Gamma_\mathrm{dark}$ is a power-independent background rate (including e.g. APD dark counts). Because of the strong NV luminescence (\cref{fig:coupling}(e),(f)), we can omit the contribution of fiber autoluminescence and fit  \cref{eq:saturation} to the data in \cref{fig:coupling}(e). We extract $\eta= 5.0(2)\times 10^{-10}$.
Writing $\eta=\eta_\mathrm{ND}\eta_\mathrm{D}\eta_\mathrm{nf}$, where $\eta_\mathrm{ND}=1.6\times 10^{-7}$ is the neutral-density (ND) filtering factor and $\eta_\mathrm{D} = 3.5\times 10^{-2}$ is the optical efficiency of the other parts of our setup (characterized separately, see SI), we extract the photon coupling efficiency at the fiber-nanobeam interface $\eta_\mathrm{nf} = 8.6(4)$\%. We note that the relatively small error here derives from the fit uncertainty of $\eta$. However, systematic uncertainties, such as the potential influence of two-photon-induced ionization of the NV centers to the neutral charge state, which would affect the detected photon rate due to the different spectrum of NV$^0$ centers \cite{Aslam2013Photo-inducedDetection,Siyushev2019PhotoelectricalDiamond} are likely to play a more important role. We therefore use a second approach to estimate $\eta_\mathrm{nf}$. 

In the second approach, we estimate $\eta_\mathrm{nf}$ from the detected NV photoluminescence using a literature value for the NV's absorption cross section $\sigma_\mathrm{NV} = 3.1(8) \times 10^{-21}$ m$^2$  for 532 nm laser excitation \cite{Wee2007Two-photonDiamond}. Far below optical saturation, the detected NV photoluminescence is given by (see SI for detailed explanation)
\begin{equation}
\label{eq:saturation2}
    \Gamma = \eta_\mathrm{f}\eta_\mathrm{D} \eta_{\mathrm{nf}}^2 \frac{\sigma_\mathrm{NV}N_\mathrm{mode}}{\sigma_\mathrm{mode}}\frac{P}{\hbar \omega}
\end{equation}
where $\sigma_\mathrm{mode}$ is the cross-sectional area of the optical mode in the nanobeam, $N_\mathrm{mode}$ is the number of NV centers within the optical mode volume, $\eta_\mathrm{f}$ is the coupling efficiency of the excitation laser into the fiber, $\omega$ is the frequency of our 532 nm laser and $\hbar$ is Planck's constant. Furthermore, we assumed that the coupling of the green laser from the fiber into the nanobeam is also given by $\eta_\mathrm{nf}$. In contrast with the first approach, this approach does not require saturating the NV photoluminescence response and can thus be conducted at very low (nW) laser power. This reduces the potential influence of two-photon-induced ionization of the NV centers to the neutral charge state \cite{Aslam2013Photo-inducedDetection,Siyushev2019PhotoelectricalDiamond}. Assuming the mode is perfectly confined within the nanobeam, we take $N_\mathrm{mode}/\sigma_\mathrm{mode}\approx N/\sigma_\mathrm{beam}$ where $\sigma_\mathrm{beam}=\SI{0.25}{\square\micro\meter}$ is the cross-sectional area of the nanobeam. From the measured $\Gamma = \SI{1.1e6}{\per\second}$ at $P=\SI{30}{nW}$, we extract $\eta_\mathrm{nf} = 14(4)$\%, similar to the value found using the first approach.  

We expect the found values for $\eta_\mathrm{nf}$ to be conservative estimates of the nanobeam-fiber coupling efficiencies due to the assumptions that all NV-emitted photons are radiated into the beam and towards the nanobeam-fiber interface, and because our two-level model neglects the non-radiative decay path via the singlet state that reduces the total photon emission rate \cite{robledo2011spin}. Compared to the state-of-the-art $\eta_\mathrm{nf}>90\%$ reported in Refs. \citenum{Burek2017Fiber-coupledInterface,Groblacher2013HighlyCavity} for single-wavelength readout, an important difference in our device is the wide-band spectrum (bandwidth $\sim\SI{200}{\nm}$) of the collected NV photoluminescence. Also, the precise alignment of the fiber tip required to optimize the coupling efficiency is affected by the abrupt motion of the nanobeam when the tether breaks; From a through-fiber measurement of the NV photoluminescence of a beam that is still attached to the bulk diamond, we estimate using the absorption cross-section that $\eta_\mathrm{nf,0} = 31(8)$\% before breaking off the beam. Other factors that reduce the efficiency include the roughness on the sidewalls and bottom side of the nanobeams, which can be improved by optimizing the fabrication process, for instance by ion-based polishing of the nanobeam sidewall \cite{Mi2019Non-contactEtching} or improved diamond etching. 

Considering the above, our estimated nanobeam-fiber coupling efficiency is remarkably high, paving the way for high-efficiency, ensemble-based NV sensing. However, the fiber autoluminescence would still exceed the single-NV photoluminescence by about an order of magnitude even in the limit $\eta_\mathrm{nf}\rightarrow 1$ (according to \cref{eq:saturation2}, $\Gamma_\mathrm{NV,max}/N=\SI{4.1e2}{\per\second\per\micro\watt}$). This indicates that single-NV sensing will only be possible if the fiber autoluminescence can be reduced, for example by incorporating hollow-core photonic crystal fibers \cite{Fujii2011ADots} that produce less fluorescence.

\begin{figure}[ht!]
    \centering
    \includegraphics[width=\textwidth]{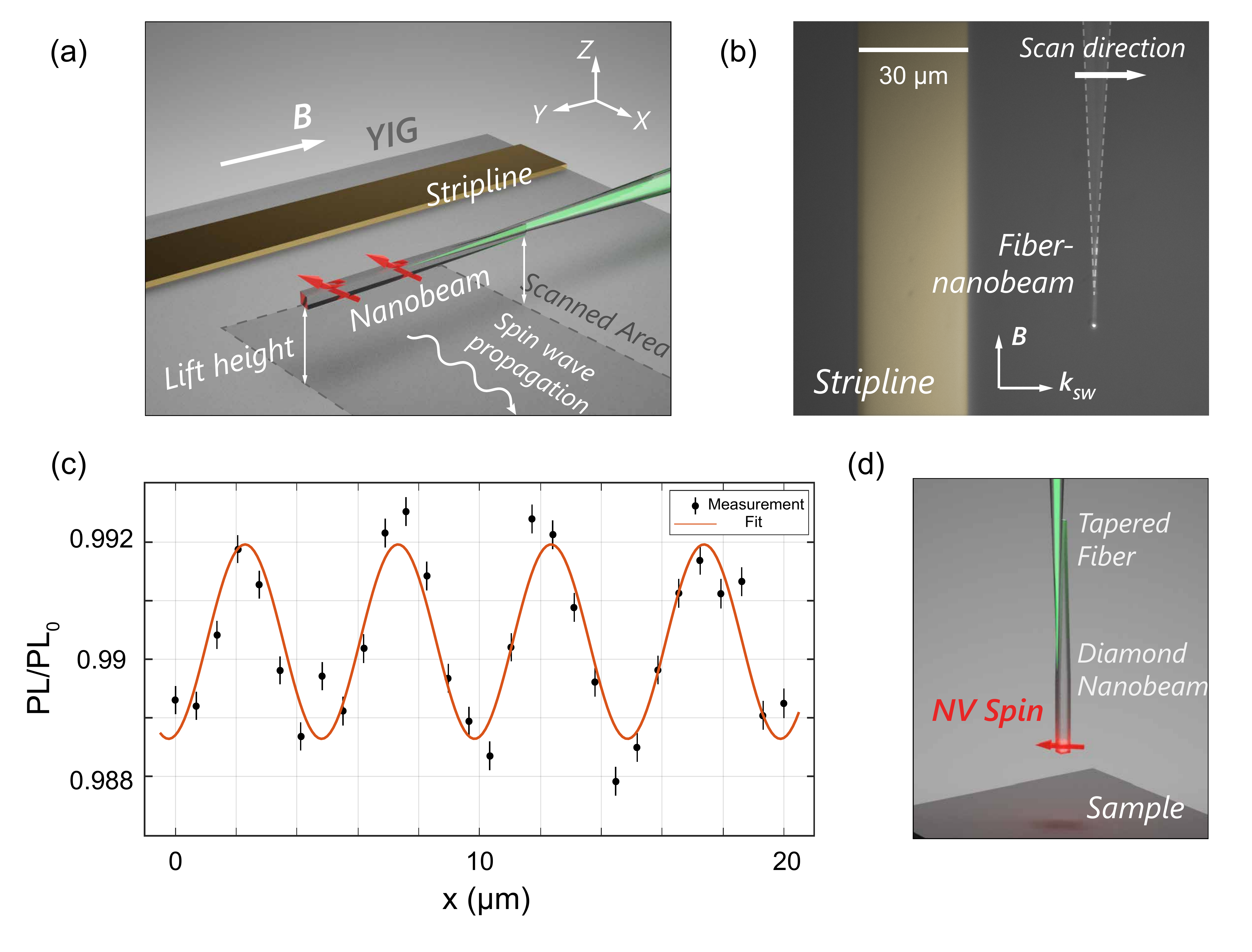}
    \caption{Proof-of-principle scanning NV magnetometry with a fiber-coupled diamond nanobeam. (a) Experimental geometry:  A fiber-coupled NV nanobeam is used to image spin waves in a 250-nm-thick film of yttrium iron garnet. The YIG-nanobeam distance is $\sim\SI{5}{\micro\meter}$. The beam is scanned perpendicularly to the beam axis. The spin waves are excited  by a microwave current in a gold stripline ($\SI{3}{\milli\meter} \times \SI{30}{\micro\meter}$). An auxiliary wire drawn across the chip (not shown) provides a reference field that interferes with the spin-wave stray field, creating a standing-wave pattern in the microwave magnetic-field amplitude \cite{Bertelli2020MagneticInsulator}. A static magnetic field $B$ is applied along the beam direction. (b) Microscope image of the experimental geometry. (c) Scanning-nanobeam imaging of a spin wave. The NV ESR contrast PL/PL$_0$ is measured by switching on and off the microwave drive at the NV ESR frequency. The error bars are estimated from assuming shot noise of the NV photoluminescence during the measurement time. A sinusoidal fit (orange) gives the measured wavelength $\lambda = \SI{5.0(1)}{\micro\meter}$. (d) Envisioned single-NV scanning diamond nanobeam for high-resolution 2D imaging.}
    \label{fig:ESR}
\end{figure}

To demonstrate the imaging capability of our fiber-coupled nanobeam sensor, we use it to image spin waves -- the wave-like excitations of spins in a magnetic material \cite{Chumak2015MagnonSpintronics} -- in a $\sim\SI{250}{\nm}$-thick film of yttrium iron garnet (YIG) \cite{Serga2010YIGMagnonics}. We excite the spin wave by sending a microwave current through a gold stripline on the YIG (\cref{fig:ESR}(a)) under a static external magnetic field $B=\SI{22}{\milli\tesla}$. The spin wave generates a microwave magnetic stray field that drives the NV spins when its frequency matches the NV ESR frequency. To create a spatial standing-wave pattern in the microwave field that we can image via the NV ESR contrast (SI), we apply an additional, spatially homogeneous reference field of the same microwave frequency using a wire above the chip \cite{Simon2021DirectionalWaves,Bertelli2020MagneticInsulator}. We scan the beam parallel to the sample surface and perpendicularly to the beam axis (\cref{fig:ESR}(a)(b)), and measure the NV ESR contrast by switching on and off the microwave drive at the ESR frequency $f = \SI{2.439}{GHz}$. The result shown in \cref{fig:ESR}(c) images the spin-wavefront in 1D with a resolution limited by the beam width and beam-sample distance. The observed wavelength of $\lambda = \SI{5.0}{\micro\meter}$ corresponds reasonably well with the $\SI{6}{\micro\meter}$ expected from the spin-wave dispersion (SI), given the uncertainty in the angle of the applied magnetic field.

\section{Conclusion}

To conclude, we demonstrated a new fiber-based approach for scanning NV magnetometry measurements. Using quasi-isotropic etching, we nano-fabricate diamond nanobeams out of single-crystal bulk diamond and couple them to tapered optical fibers. We read out ensemble NV signals through the fiber-nanobeam coupling with an estimated efficiency of 8.6(4)\% at the coupling interface. As a demonstration, we show that our device can function as a scanning sensor to measure in 1D the planar spin wave in YIG.

A remaining challenge lies in increasing the control over the angle and position when attaching the nanobeam to the fiber. While we found that we can consistently break off the beams and attach them to a fiber, their relative position after the breaking process is not entirely under control due to the abrupt motion of the fiber-nanobeam when the tether breaks. We expect that reducing the tether width further, or reducing the nanobeam surface roughness via improved etching or ion-based polishing \cite{Mi2019Non-contactEtching}, could improve the coupling efficiency. Additionally, we found that transporting the fiber-nanobeam probe is challenging due to vibration and/or static electricity that cause the nanobeam to detach. Possible solutions could include coating the nanobeam/fiber to enhance the fiber-diamond adhesion, and transporting the device in a electrostatic-free environment such as a metal enclosure. Further steps towards 2D magnetic imaging (\cref{fig:ESR}(d)) include deterministically placing NV centers at the end of the nanobeams by e.g. pre-localizing NV centers \cite{Wan2018EfficientReflector} or deterministic implantation \cite{Schukraft2016PrecisionWaveguides}. With above mentioned efforts, our work holds potential for implementation in low-temperature setups with reduced heat load and easier alignment, opening another possibility for imaging weak magnetic effects at low temperature e.g. currents in quantum Hall devices  \cite{Uri2020NanoscaleGraphene} and Josephson junctions \cite{Roditchev2015DirectCores}.

\section{Methods}

\subsection*{Magnetometry with NV centers}
The NV center is a spin-1 system. The ground state of an NV center splits into three spin substates $m_S=0,\pm1$, and the microwave-driven transition between $m_S=0$ and $m_S=\pm1$ states can be detected via the photoluminescence intensity under non-resonant green laser excitation: Once the frequency of applied microwave matches the $m_S=0\rightarrow\pm1$ transition frequencies (ESR frequencies), the photoluminescence emission of the NV center will decrease due to the higher non-radiative decay rate of the $m_S=\pm1$ states. Applying an external magnetic field lifts the degeneracy of the $m_S=\pm1$ states, allowing magnetic field measurement through measuring the ESR frequencies. More detailed information on the working principle of NV centers can be found in Refs. \citenum{Schirhagl2014Nitrogen-vacancyBiology,Rondin2014MagnetometryDiamond,Doherty2013TheDiamond,Casola2018ProbingDiamond}. 

\subsection*{Device Fabrication}

We fabricate the diamond nanobeams on single crystal CVD diamond (Element-six DNV-B14) with ensemble NV centers generated during growth. Before fabrication, we mechanically polish the diamond surface down to $R_a\sim\SI{2}{nm}$ (Almax EasyLab) and clean the diamond chip with fuming nitric acid ($\mathrm{HNO_3}$).

To fabricate the beams, we first deposit a 200 nm layer of $\mathrm{Si_3N_4}$ on the surface with PECVD ($\SI{20}{sccm}$ $\mathrm{SiH_4}$/$\SI{20}{sccm}$ $\mathrm{NH_3}$/$\SI{980}{sccm}$ $\mathrm{N_2}$, deposited at $\SI{300}{\degreeCelsius}$, Oxford Instruments Plasmalab 80 Plus) as the hard mask. We then spin-coat a $\sim\SI{400}{nm}$ layer of e-beam resist (AR-P 6200-13) and a $\sim\SI{30}{nm}$ conductive layer of Elektra-92 on top to write the pattern with e-beam lithography (Raith EBPG5200). We transfer the e-beam pattern from the resist to the SiN hard mask by an anisotropic ICP-RIE etch with CHF$_3$/O$_2$ ($\SI{60}{sccm}/\SI{6}{sccm}$, $\SI{50}{W}$ RF and $\SI{500}{W}$ ICP at $\SI{20}{\degreeCelsius}$, AMS 100 I-speeder). We then remove the resist with dimethylformamide (DMF) and subsequent Piranha cleaning (96\% H$_2$SO$_4$ and 31\% H$_2$O$_2$, 3:1 mixed at $\SI{80}{\degreeCelsius}$). An ICP-RIE etch with O$_2$ ($\SI{50}{sccm}$, $\SI{90}{W}$ RF and $\SI{1100}{W}$ ICP at $\SI{20}{\degreeCelsius}$, Oxford Instruments Plasmalab 100) transfers the pattern onto the diamond. 

To protect the sidewalls of the structure during the subsequent undercut etch, we deposit $\sim\SI{20}{nm}$ of $\mathrm{Al_2O_3}$ with ALD (280 cycles, at $\SI{105}{\degreeCelsius}$, Oxford Instruments FlexAL)  and remove the $\mathrm{Al_2O_3}$ on the topside with another ICP-RIE with BCl$_3$/Cl$_2$ ($\SI{45}{sccm}$/$\SI{5}{sccm}$, $\SI{10}{W}$ RF and $\SI{600}{W}$ ICP at $\SI{20}{\degreeCelsius}$, Oxford Instruments Plasmalab 100). We do the final undercut of the beams with quasi-isotropic O$_2$ ICP-RIE at $\SI{65}{\degreeCelsius}$ ($\SI{50}{sccm}$, $\SI{0}{W}$ RF and $\SI{2500}{W}$ ICP, Oxford Instruments Plasmalab 100). For the sample discussed in the main text, completely undercutting the $\sim\SI{500}{nm}$ wide nanobeams took 12 hours. The etch rate of our quasi-isotropic etching process is mainly limited by the maximum etching temperature of our system, and varies over different design, sample and etcher condition. All the masks are eventually cleaned with 40\% hydrofluoric acid (HF, 10 min). Further details of the relevant recipe parameters can be found in Ref. \citenum{max2021thesis}.

We fabricate the tapered fibers by wet etching commercial optical fibers (S630-HP) with 40\% HF. We dip one end of the fiber into the acid and pull it out at constant speed using a motorized translation stage (Thorlabs MTS25-Z8). The tapering angle can thus be controlled by tuning the pulling speed \cite{Burek2017Fiber-coupledInterface}.



\section{Author Information}

\subsection{Funding}
This work is supported by the Dutch Science Council (NWO) through the NWA grant 1160.18.208 and the Kavli Institute of Nanoscience Delft. N.C. acknowledges support from the joint research program “Modular quantum computers” by Fujitsu Limited and Delft University of Technology, co-funded by the Netherlands Enterprise Agency under project number PPS2007.

\subsection{Author Contributions}
 Y.L., R.N. and T.v.d.S. conceived the experiments. Y.L. and F.G. developed the measurement setup and performed the experiments. Y.L., R.N., N.C., S.G. and R.H. developed the diamond fabrication recipes and Y.L. fabricated the diamond nanobeam samples. S.K. prepared the YIG sample for the imaging measurements. Y.L. and T.v.d.S. analyzed the results. Y.L. and T.v.d.S. wrote the manuscript with contributions from all coauthors.

 \subsection{Notes}
  The authors declare no competing financial interest.

 \subsection{Data availability}
 All data plotted in the figures are this work are available at zenodo.org with identifier 10.5281/zenodo.7561825. Additional data related to this paper are available upon request.

\begin{acknowledgement}

The authors thank T. Bredewoud for theoretical simulations of the fiber-nanobeam coupling, B.G. Simon, M. Ruf and C. van Egmond for the help in the fabrication process.

\end{acknowledgement}

\begin{suppinfo}

Estimation of coupling efficiency using the absorption cross section method; Explanation on the spin wave dispersion in YIG, and the principles of using NV centers to image propagating spin waves.

\end{suppinfo}

\bibliography{achemso-demo}

\end{document}


\pagebreak

\section{Estimating the Collection Efficiency through Absorption Cross Section}

Consider a single NV center inside a diamond nanobeam. If the area of the optical mode cross section inside the beam is approximately the cross section of the nanobeam itself $A_\mathrm{beam}$, then the probability for an NV center to absorb a single photon and trigger an excitation will be

\begin{equation}
    \mathcal{P} = \frac{\sigma}{A_\mathrm{beam}}
\end{equation}
where $\sigma = \SI{3.1(8)e-21}{m^2}$ is the absorption cross section of the NV center\cite{Wee2007Two-photonDiamond}. Therefore when a total of $N_\mathrm{ph}$ photons enter a beam containing $N_\mathrm{NV}$ NVs, the total number of excitations (thus total number of emitted photons) will be

\begin{equation}
    N_\mathrm{PL} = \frac{N_\mathrm{ph}N_\mathrm{NV}\sigma}{A_\mathrm{beam}}.
\end{equation}
Note that the effect of saturation discussed in the main text is not taken into account here, thus the optical power needs to stay well below saturation for this equation to hold.

Now consider a power of $P$ being sent into the fiber with a fiber-coupling efficiency of $\eta_\mathrm{f}$. The rate of excitation photons that eventually end up in the beam mode will then be

\begin{equation}
    \Gamma_\mathrm{exc} = \frac{P}{\hbar\omega}\eta_\mathrm{f}\eta_\mathrm{nf}
\end{equation}
where $\eta_\mathrm{nf}$ is the coupling efficiency at the fiber-nanobeam interface. Thus the photoluminescence rate will be
\begin{equation}
    \Gamma_\mathrm{PL} = \frac{P}{\hbar\omega}\frac{N_\mathrm{NV}\sigma}{A_\mathrm{beam}}\eta_\mathrm{f}\eta_\mathrm{nf}
\end{equation}
and the measured photon rate is (assuming coupling efficiency at the fiber-nanobeam interface is equal for both directions)
\begin{equation}
    \label{eq:rate}
    \Gamma_\mathrm{meas} = \frac{P}{\hbar\omega}\frac{N_\mathrm{NV}\sigma}{A_\mathrm{beam}}\eta_\mathrm{f}\eta^2\eta_\mathrm{D}
\end{equation}
where $\eta_\mathrm{D}$ is the fraction of photons exiting the fiber that are eventually detected by the APD.

\begin{figure}[ht!]
    \centering
    \includegraphics[width = 0.8\textwidth]{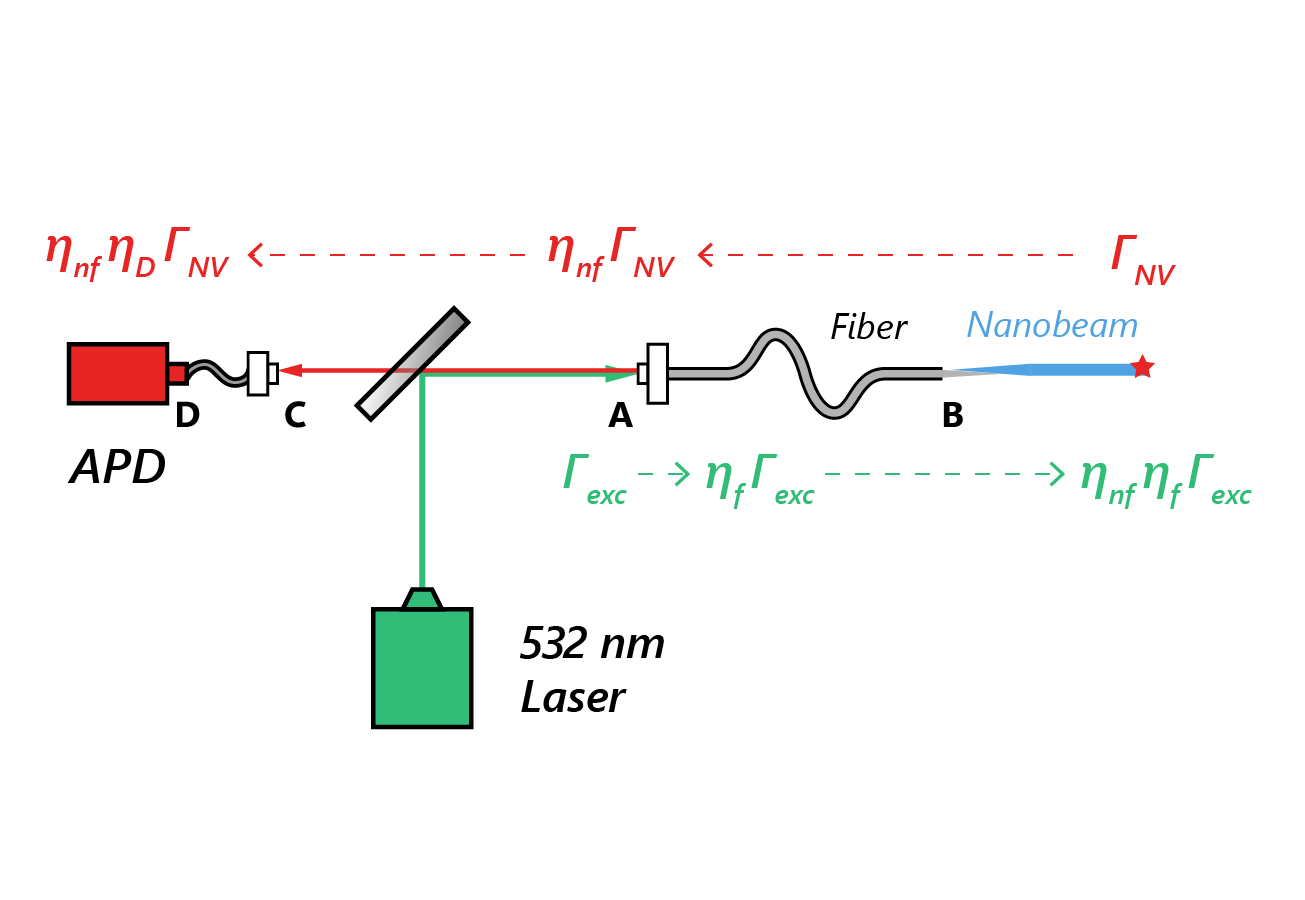}
    \caption{Overview of the efficiencies involved in our measurement. $\Gamma_\mathrm{exc}$: Photon rate of excitation laser. $\eta_\mathrm{f}$: Efficiency of free-space excitation laser coupling into the fiber. $\eta_\mathrm{nf}$: Efficiency of adiabatic light coupling at the fiber-nanobeam interface. $\Gamma_\mathrm{NV}$: Photoluminescence rate of NV centers inside the nanobeam. $\eta_\mathrm{D}$: fraction of total photons exiting the fiber that are eventually detected by the APD, consisting of losses in free space optics (e.g. on mirrors and filters), fiber coupling efficiency at the APD input and the detection efficiency of the APD.}
    \label{fig:efficiency}
\end{figure}

Therefore from \cref{eq:rate} one can estimate $\eta_\mathrm{nf}$ from $P$ and $\mathcal{R}_\mathrm{meas}$, both of which are measured experimentally. For the device mentioned in the main text, we measured $\Gamma_\mathrm{meas} = \SI{1.08e6}{\per\second}$ at $P = \SI{30}{\nano\watt}$. For the efficiencies, \cref{fig:efficiency} shows a simplified overview of the measurement setup and the efficiencies involved. To experimentally determine $\eta_\mathrm{f}$, we connect a non-tapered fiber (same model as the tapered one) to the fiber coupler (A), send in a green laser with power $P_\mathrm{in}$ measured in front of the coupler, measure the power at the output (B) of the fiber $P_\mathrm{out}$ and determine $\eta_\mathrm{f}=P_\mathrm{out}/P_\mathrm{in}\approx 0.35$. For $\eta_\mathrm{D}$, we send in a red laser from B, measure the power at fiber output (A) and the APD fiber input (C and D), and determine $\eta_\mathrm{D}\approx 0.035$ (including the free-space trasmission rate $\eta_\mathrm{A\rightarrow C}\approx0.5$, APD fiber coupling efficiency $\eta_\mathrm{C\rightarrow D}\approx0.1$ and APD detection efficiency $\eta_\mathrm{APD}=0.7$). Subtituting all the numbers into \cref{eq:rate}, we can calculate the coupling efficiency at the interface
\begin{equation}
    \eta_\mathrm{nf} = 14(4)\%.
\end{equation}

\section{Imaging the Spin Waves in YIG with NV centers}

Spin waves can be excited in the YIG thin film with the AC magnetic field we apply through the stripline. To model the spin dynamics, we consider our YIG layer as an infitite film parallel to the $xy$ plane, infinite in $x$ and $y$, and has a finite thickness $t$ in $z$ direction. We can calculate the dispersion of the spin waves using the Landau-Lifshitz-Gilbert (LLG) equation which governs the dynamics of magnetization:
\begin{equation}
    \mathbf{\dot{m}} = -\gamma\mathbf{m}\times(\mathbf{B}_\mathrm{eff}+\mathbf{B}_\mathrm{AC})-\alpha\mathbf{\dot{m}}\times\mathbf{m}
\end{equation}
in which $\mathbf{m}(\mathbf{r})$ is the unit vector in the direction of the magnetization $\mathbf{M}(\mathbf{r})=M_s\mathbf{m}(\mathbf{r})$ (with $M_s$ being the saturation magnetization), $\mathbf{B}_\mathrm{AC}$ is the AC driving field, $\gamma$ is the gyromagtic ratio and $\alpha$ is the Gilbert damping. The effective field $\mathbf{B}_\mathrm{eff}$ takes into account the static external magnetic field $\mathbf{B}_0$, the demagnetizing field $\mathbf{B}_d$ and the exchange interaction.

Since spin waves are propagating waves of spin precession around the magnetization axis in equilibrium, we are interested in the dynamics of the \emph{transverse} magnetization
\begin{equation}
    \mathbf{m}_\perp = \chi\mathbf{B}_\mathrm{AC,\perp}
\end{equation}
where $\mathbf{m}_\perp = m_{x'}\hat{\mathbf{x'}}+m_{y'}\hat{\mathbf{y'}}$ is the transverse components of the magnetization vector, with the magnetic frame $(x',y',z')$ defined such that $\mathbf{m}(\mathbf{r})=\hat{\mathbf{z'}}$ in equilibrium. $\mathbf{B}_\mathrm{AC,\perp} = \mathbf{B}_\mathrm{AC,x'}\hat{\mathbf{x'}}+\mathbf{B}_\mathrm{AC,y'}\hat{\mathbf{y'}}$ is the transverse drive field in the magnetic frame, and $\chi$ is the \emph{transverse susceptibility}. Following the formalism by Rustagi et al.\cite{Rustagi2020SensingRelaxometry}, we can find $\chi$ by solving the LLG equation in $k$-space, and determine the spin wave dispersion $\omega_\mathrm{sw}(\mathbf{k})$ by finding the singularity of $\chi$, which eventually gives us:
\begin{equation}
    \omega_\mathrm{sw}(\mathbf{k})=\sqrt{\omega_2\omega_3-\omega_1^2},
    \label{eq:swdisp}
\end{equation}
where $\mathbf{k}=(k_x,k_y)$ is the in-plane wave vector. Parameters $\omega_1\sim\omega_3$ are in general functions of the angle between the $z$ and $z'$ axes, the direction of $\mathbf{k}$, effective static field and the film thickness\cite{Rustagi2020SensingRelaxometry}. In our measurements, we apply the external magnetic field in-plane, and we excite and detect spin waves in the Damon-Eshbach (DE) regime where the spin wave wave vector $\mathbf{k}$ is perpendicular to the external magnetic field $\mathbf{B}_0$. Under these conditions, $\omega_1\sim\omega_3$ are reduced to:
\begin{align}
    \omega_1 &= 0;\\
    \omega_2 &= \omega_B + \omega_Dk^2 + \omega_M(1-f);\\
    \omega_3 &= \omega_B + \omega_Dk^2 + \omega_Mf.
\end{align}
where we defined $\omega_B=\gamma B_0$, $\omega_D=\gamma D/M_s$ with $D$ being the spin stiffness that characterizes the exchange interaction between spins, $\omega_M=\gamma\mu_0M_s$, $f=1-(1-e^{-kt})/kt$ with $t$ being the film thickness.

The spin waves generate an AC magnetic field at the spin wave frequency $\omega_\mathrm{sw}$, and the amplitude of the field varies in space according to the wave number $k$. In our system we use a stripline to excite spin waves traveling in $x$-direction with a planar wave front, which can be written as 
\begin{equation}
    B_\mathrm{sw}(x,t)=B_\mathrm{sw,0}e^{i(kx+\omega_\mathrm{sw}t)}.
\end{equation}
This field can then drive the NV-ESR transitions when the spin wave frequency matches the ESR frequency. In order to image the spatial variation of $B_\mathrm{sw}$, we apply a homogeneous reference field with the same frequency $B_\mathrm{ref}(t) = B_\mathrm{ref,0}e^{i\omega_\mathrm{sw}t}$ through a bonding wire located above the sample. This creates a spatial variation in the field amplitude at the spin wave wavelength:
\begin{equation}
    |B_\mathrm{sw}(x,t)+B_\mathrm{ref}(t)|^2 = B^2_\mathrm{sw,0}+B^2_\mathrm{ref,0}+2B_\mathrm{sw,0}B_\mathrm{ref,0}\cos{kx}.
\end{equation}
This results in a spatially varying Rabi frequency of the NV-ESR transition
\begin{equation}
    \Omega_R(x) = \frac{\gamma}{\sqrt{2}}|B_\mathrm{sw}(x,t)+B_\mathrm{ref}(t)|
\end{equation}
which directly determines the NV-ESR contrast $\mathcal{C}=(\mathrm{PL}-\mathrm{PL_0})/\mathrm{PL_0}$ (defined in the main text) through the relation\cite{Dreau2011AvoidingSensitivity}
\begin{equation}
    \mathcal{C}(x)\propto\frac{\Omega_R^2(x)}{\Omega_R^2(x)+\Delta}
\end{equation}
where the parameter $\Delta$ is determined by the optical excitation power of NV centers, therefore remains constant in our measurement scheme. As a result, the measured ESR contrast along the spin wave propagation direction has the same periodicity as the spin wave itself, allowing the imaging of spin wave through measuring ESR contrast. 

Furthermore, from the ESR frequency one can determine both the detected spin wave frequency (as they should be exactly the same) and the external magnetic field strength, and theoretically determine the wavelength of the measured spin wave through \cref{eq:swdisp}. This gives us $\lambda\sim\SI{6}{\micro\meter}$ for our measurement in fig.4 of the main text, which agrees reasonably well with our measurement given the imperfect alignment of both the nanobeam and the external magnetic field.

\bibliography{achemso-demo-supp}